\documentstyle[prl,aps,floats,psfig]{revtex}

\begin{document}

\widetext
\draft
\twocolumn[\hsize\textwidth\columnwidth\hsize\csname @twocolumnfalse\endcsname
\title{Reactive Turbulent Flow in Low-Dimensional, Disordered Media}

\author{
Michael W. Deem$^1$
and 
Jeong-Man Park$^{1,2}$ 
}
\address{
$^1$Chemical Engineering Department, University of
California, Los Angeles, CA  90095-1592\\
$^2$Department of Physics, The Catholic University, Seoul, Korea}

\maketitle

\begin{abstract}
We analyze the reactions $A+A \to \emptyset$ and $A + B \to
\emptyset$ occurring in a model of turbulent flow in two dimensions.  
We find the reactant concentrations at long times, 
using a field-theoretic renormalization
group analysis.
We find a variety of 
interesting behavior, including,
in the presence of potential disorder,
 decay rates faster than that for 
well-mixed reactions.
\end{abstract}

]

\section{Introduction}

The behavior of chemical reactions in turbulent flow determines
how certain types of chemical reactors function, how combustion
occurs in engines, how 
smog  is produced in the atmosphere \cite{Bilger}, and how certain types of
planktonic predators feed in the ocean \cite{Kiorboe}.
Reactive turbulent flow is usually analyzed by continuum
reaction/transport equations for the reactants \cite{Bilger,Hill}.
We now know, however, that the upper critical dimension in which
such mean-field equations fail is two \cite{Peliti,Lee2,Deem1,Deem2}.
We provide here the first renormalization group, field-theoretic treatment of
reactive turbulent flow in two dimensions.

We consider a low concentration of reactants immersed in an
isotropic, turbulent
fluid flow.  The interplay of reaction, turbulent mixing, and
trapping by disorder will be shown to lead to novel kinetics
at long times.  So as to access the most interesting regime, we
will consider a two-dimensional system.    At low reactant
concentrations, the dynamics of the fluid will not be affected by
the kinetics of the reaction.  As such, the effect of the turbulence is
simply to advect and to mix the reactants.

Our intention is not to derive a theory of turbulence but rather to
derive a theory of bimolecular kinetics in the presence of isotropic turbulence
and potential disorder.  We, therefore, employ the same statistical theory
of turbulence conventionally used to study turbulent transport of passive
scalars \cite{Fisher,Kravtsov1}.  For a review of this approach, see
\cite{Bouchaud3}.
We assume, in particular, that the turbulent fluid which advects the
reactants can be modeled as a quenched, random, Gaussian velocity field
with the correct statistics.
  While the fluid velocity streamlines produced
by this conventional approach do not satisfy the Navier-Stokes equations, the
correct transport properties of the reactants are captured.  The
correct Kolmogorov energy cascade and Richardson separation laws,
for example, are produced.   One can imagine using
a more detailed model of the fluid
mechanics.
Avellaneda and Majda have, for example, used statistical flow fields that
depend on both space and time to model turbulent transport
\cite{Majda1,Majda2,Majda3}.  On an even more detailed level, one could
use statistical flow fields that satisfy the Navier-Stokes equations.
Renormalization group theories for 
flow fields of this type have been derived by
Forster, Nelson, and Stephen \cite{Forster} and later by
Yakhot, Orszag, and coworkers \cite{Orszag1,Orszag2,Orszag3}.  We
settle for the simplest description of the fluid mechanics that
 captures the essence of turbulent transport.  Our results should
not be sensitive to this assumption, since the physical processes that
appear to dominate the long-time kinetics depend only on the
overall transport properties of the fluid.
In support of the generality of our results, we note that 
Deering and West have  used mean field theory to analyze 
the reaction $A + B \to \emptyset$
in a time-dependent, but spatially uncorrelated, model of 
 isotropic turbulence.  Where mean field theory is expected to
work (where the renormalized reaction rate does not enter
the prediction for the
long-time reactant concentration) their results agree with ours.

In this article, we use a field-theoretic approach to analyze
reaction kinetics in a model of turbulent fluid flow.
A definition of the fluid flow and the reaction
kinetics
is given in Section II.  The reactive turbulence problem is mapped to a
field theory that is convenient for analysis in Section III.  The 
long-time behavior of the $A + A \to \emptyset$ reaction is 
derived by a renormalization group analysis in Section IV.
The long-time behavior of the $A + B \to \emptyset$ reaction is 
derived in a similar fashion in Section V.  We conclude with
a comparison to experimental results in Section VI.

\section{Definition of Reactive Turbulent Flow}
In our formulation, an isolated reactant undergoes biased Brownian
motion in the fluid streamlines, reacting at a given rate with
other nearby reactants.  In the absence of reaction, the motion of
the particles can be described by a Langevin equation:
\begin{equation}
\frac{d {\bf x}_i}{d t} = \beta D {\bf F}({\bf x}_i) + 
{\mbox{ \boldmath $\eta$}}(t) \ ,
\label{1}
\end{equation}
where the inverse temperature is given by $\beta = 1/(k_B T)$, and
$D$ is the diffusivity.
Here the position of particle $i$, ${\bf x}_i$, undergoes  advection
due to  forces 
from the fluid flow
and diffusion
due to forces
from the random, thermal motion of the fluid.  The random, thermal 
noise has
a correlation determined by the diffusion coefficient:
\begin{equation}
\langle \eta_\mu (t) \eta_\nu(t') \rangle = 2 D \delta_{\mu \nu}
\delta(t-t') \ .
\label{2}
\end{equation}
  We choose the forces coming from the fluid stream lines
so that they mimic turbulence.  Defining
\begin{equation}
\langle F_\mu({\bf x}) F_\nu({\bf x}') \rangle =
   G_{\mu \nu}({\bf x}-{\bf x}') \ ,
\label{3}
\end{equation}
we choose
\begin{equation}
\hat G_{\mu \nu} ({\bf k}) = \hat \chi_{\phi \phi}({\bf k}) \left( \delta_{\mu \nu} k^2 - 
        k_\mu k_\nu \right)
  + \hat \chi_{uu}({\bf k}) k_\mu k_\nu \ ,
\label{4}
\end{equation}
where the Fourier transform of the correlation function is
$\hat G_{\mu \nu} ({\bf k}) = \int d^d {\bf x} G_{\mu \nu}({\bf x})
\exp(i {\bf k} \cdot {\bf x})$ in $d$ dimensions.
Here $\hat \chi_{\phi \phi}({\bf k})$ is the correlation function of the
stream function that gives rise to the turbulent fluid flow, and
$\hat \chi_{uu}({\bf k})$ is the correlation function of a quenched, random
potential, $u({\bf x})$, which we have included for generality.
To mimic turbulence, we choose
\begin{eqnarray}
\hat \chi_{\phi \phi}({\bf k}) &= & \frac{\sigma}{k^{2+y}}
\nonumber \\
\hat \chi_{uu}({\bf k}) &=& \frac{\gamma}{k^{2+y}} \ .
\label{5}
\end{eqnarray}
Isotropic turbulence is modeled by $y = 8/3$ and $\gamma = 0$.

Of course, we are interested in reactive, turbulent flow.  We consider
two different reactions:
\begin{equation}
A+A
~{\mathrel{\mathop{\to}\limits^{\lambda_0}}}~ \emptyset
\label{6}
\end{equation}
and
\begin{equation}
A+B
~{\mathrel{\mathop{\to}\limits^{\lambda_0}}}~ \emptyset \ .
\label{7}
\end{equation}
Here $\lambda_0$ is the conventional reaction rate.  
We place our reactants on a square lattice, of lattice spacing $h$.
This lattice spacing implies a cutoff in Fourier space of
$\Lambda = 2 \pi / h$.
Reaction
occurs between two particles, at rate $\lambda_0/h^2$, only when they
are on the same lattice site.  The diffusion and advection occurs on this
same lattice.

\section{Field-Theoretic Representation}
The quantity of interest is the long-time concentration of the
reactants.
The presence of the quenched fluid stream lines and the
quenched, external potential makes direct analysis of the dynamics
rather difficult.  Perturbation theory fails due to singularities
in the forces at small $k$.  We, therefore, map the above description
onto a field theory and analyze the field theory using renormalization
group theory.  We assume that the concentration of reactants
is initially Poisson, with average density $n_0$.

A field theory is derived
by identifying a master equation, writing the master equation
in terms of creation and annihilation operators, and
using the coherent state representation \cite{Peliti,Lee1}.  The
random potential is incorporated with the replica trick \cite{Kravtsov1},
using $N$ replicas of the original problem.
For reaction (\ref{6}), 
the concentration of A at time $t$, averaged over the
initial conditions, $c_A({\bf x},t)$, is given by
\begin{equation}
c_A({\bf x},t) = \lim_{N \to 0} \langle a({\bf x},t) \rangle \ ,
\label{8}
\end{equation}
where the average is taken with respect to  $\exp(-S_{AA})$, 
\begin{eqnarray}
S_{AA} &=& \int d^d {\bf x} \int_0^{t_f} d t
 \bar a_\alpha({\bf x},t) \left[
\partial_t - D \nabla^2 + \delta(t)
 \right]
 a_\alpha({\bf x},t)
 \nonumber \\ 
 && 
+\frac{\lambda_0}{2} \int d^d {\bf x} \int_0^{t_f} d t \bigg[
2 \bar a_\alpha({\bf x},t)
 a_\alpha^2({\bf x},t)
\nonumber \\ &&
+ \bar a_\alpha^2({\bf x},t) 
 a_\alpha^2({\bf x},t) 
\bigg]
  -n_0 \int d^d {\bf x} \bar a_\alpha({\bf x},t)
\nonumber \\ 
&& -\frac{\beta^2 D^2}{2}
\int d t_1 d t_2 \int_{{\bf k}_1 {\bf k}_2 {\bf k}_3 {\bf k}_4}
\nonumber \\ 
&& \times (2 \pi)^d \delta({\bf k}_1+{\bf k}_2+{\bf k}_3+{\bf k}_4)
\nonumber \\ &&\times 
\hat{\bar a}_{\alpha_1}({\bf k}_1, t_1)
\hat{     a}_{\alpha_1}({\bf k}_2, t_1)
\hat{\bar a}_{\alpha_2}({\bf k}_3, t_2)
\hat{     a}_{\alpha_2}({\bf k}_4, t_2)
\nonumber \\ &&\times [
{\bf k}_1 \cdot ({\bf k}_1+{\bf k}_2)
{\bf k}_3 \cdot ({\bf k}_3+{\bf k}_4)
\hat\chi_{uu}(\vert {\bf k}_1+{\bf k}_2\vert)
\nonumber \\ 
&&+ 
{\bf k}_1 \times {\bf k}_2~
{\bf k}_3 \times {\bf k}_4
\hat\chi_{\phi \phi}(\vert {\bf k}_1+{\bf k}_2\vert)]
\ .
\label{9}
\end{eqnarray}
Summation is implied over replica indices.  The notation
$\int_{\bf k}$ stands for $\int d^d {\bf k} / (2 \pi)^d$.
The upper time limit in the action is arbitrary as long as 
$t_f \ge t$.  We do not dwell on the construction of this field
theory.  It differs from that for
reaction in a random potential field only by the inclusion of the
random stream line terms \cite{Deem1}.

For distinct reactants, reaction (\ref{7}), a field theory
can also be derived.  The relevant action has the
form
\begin{eqnarray}
S_{AB} &=&
\int d^d {\bf x} \int_0^{t_f} d t\,
 \bar a_\alpha({\bf x},t) \left[
\partial_t - D \nabla^2 + \delta(t)
 \right]
 a_\alpha({\bf x},t)
 \nonumber \\
&&+
\int d^d {\bf x} \int_0^{t_f} d t\,
 \bar b_\alpha({\bf x},t) \left[
\partial_t - D \nabla^2 + \delta(t)
 \right]
 b_\alpha({\bf x},t)
 \nonumber \\
 &&+
\lambda_0 \int d^d {\bf x} \int_0^{t_f} d t \bigg[
\bar a_\alpha({\bf x},t)
 a_\alpha({\bf x},t)
 b_\alpha({\bf x},t)
\nonumber \\
&&+
 \bar b_\alpha({\bf x},t)
 a_\alpha({\bf x},t)
 b_\alpha({\bf x},t)
\nonumber \\
&&+ \bar a_\alpha({\bf x},t)
 a_\alpha({\bf x},t)
 \bar b_\alpha({\bf x},t)
 b_\alpha({\bf x},t)
\bigg]
\nonumber \\
 && -n_0 \int d^d {\bf x}
\left[ \bar a_\alpha({\bf x},0) +
\bar b_\alpha({\bf x},0) \right]
\nonumber \\
&& -\frac{\beta^2 D^2}{2}
\int d t_1 d t_2 \int_{{\bf k}_1 {\bf k}_2 {\bf k}_3 {\bf k}_4}
\nonumber \\
&& \times (2 \pi)^d \delta({\bf k}_1+{\bf k}_2+{\bf k}_3+{\bf k}_4)
\nonumber \\ &&\times
\left[
\hat{\bar a}_{\alpha_1}({\bf k}_1, t_1)
\hat{     a}_{\alpha_1}({\bf k}_2, t_1) -
\hat{\bar b}_{\alpha_2}({\bf k}_1, t_1)
\hat{     b}_{\alpha_2}({\bf k}_2, t_1)
\right]
\nonumber\\
&& \times
\left[
\hat{\bar a}_{\alpha_3}({\bf k}_3, t_2)
\hat{     a}_{\alpha_3}({\bf k}_4, t_2) -
\hat{\bar b}_{\alpha_4}({\bf k}_3, t_2)
\hat{     b}_{\alpha_4}({\bf k}_4, t_2)
\right]
\nonumber \\ &&\times [
{\bf k}_1 \cdot ({\bf k}_1+{\bf k}_2)
{\bf k}_3 \cdot ({\bf k}_3+{\bf k}_4)
\hat\chi_{uu}(\vert {\bf k}_1+{\bf k}_2\vert)
\nonumber \\
&&+
{\bf k}_1 \times {\bf k}_2~
{\bf k}_3 \times {\bf k}_4
\hat\chi_{\phi \phi}(\vert {\bf k}_1+{\bf k}_2\vert)]
\ .
\label{9a}
\end{eqnarray}
This action also differs from that for
reaction in a random potential field only by the inclusion of the
random stream line terms \cite{Deem2}.  
The concentrations averaged over initial conditions
are given by
\begin{eqnarray}
 c_A({\bf x},t)  &=&
 \lim_{N \to 0} \langle a({\bf x},t) \rangle \ 
 \nonumber \\
 c_B({\bf x},t)  &=&
 \lim_{N \to 0} \langle b({\bf x},t) \rangle \ 
\ ,
\label{10}
\end{eqnarray}
where the average on the right hand side
is taken with respect to  $\exp(-S_{AB})$.
  So as to reach
the most interesting
scaling limit, we have taken the initial average densities to be the same,
$ c_A({\bf x},0)= c_B
({\bf x},0)= n_0$. 
 For simplicity, we have also
assumed equal diffusivities of the two species, $D_A = D_B = D$.
Note that the $A$ and
$B$ particles experience an identical force due to the fluid stream lines
but an opposite force due to the quenched, external potential.

\section{The $A+A \to \emptyset$ Reaction}
Let us first consider how turbulence will
affect the $A+A \to \emptyset$ reaction 
in the absence of potential disorder.
Without turbulent mixing, this reaction is diffusion-limited.
 The concentration decays at long times
as $c_A(t) \sim \ln(t/t_0) / (8 \pi D t)$, with $t_0
\approx h^2/D$ \cite{Peliti}.  This decay is slower than the 
$c_A(t) \sim 1/(k^* t)$ that would be predicted by
simple mean-field kinetics for a well-mixed
reaction with effective reaction rate $k^*$.  
Turbulence mixes the reactants, tending to 
eliminate the transport limitation on the reaction rate.
As we will see, turbulence will cause the reactant 
concentration to follow the mean-field result, with $k^* \le \lambda_0$.

We analyze the field theory (\ref{9}) via renormalization
group theory.
The flow equations in two dimensions,
to one loop order, are
\begin{eqnarray}
\frac{d \ln n_0}{d l} &=& 2
\nonumber \\
\frac{d \ln \lambda}{d l} &=& - 
   \frac{\lambda}{4 \pi D} - g
\nonumber \\
\frac{d \ln g}{d l} &=& y - 2 g
\ ,
\label{11}
\end{eqnarray}
where the dimensionless coupling constant is given by
$g = \sigma \beta^2 \Lambda^{-y} / (4 \pi)$.
The dynamical exponent is given by
\begin{equation}
z = 2  - g \ .
\label{12}
\end{equation}
We see that the flow equations lead to a non-zero fixed point for the
coupling $g^* = y/2$.  This fixed point is probably exact, if we
assume that the reaction does not affect the transport properties
\cite{Deem1,Bouchaud3}.

We determine the long-time decay from the flow equations via matching
to short-time perturbation theory \cite{Deem1}.
The flow equations are integrated to a
time such that
$t(l^*) = t \exp[-\int_0^{l^*} z(l) dl ] = t_0$.
At short times, we
 find the mean-square displacement of an unreactive particle from
$\left\langle r^2(t(l^*), l^*) \right\rangle = 4 D t(l^*)$ and
the concentration of reactants from 
$c_A(t(l^*), l^*) = 1/[1/n_0(l^*) + \lambda(l^*) t(l^*)]$.
The long-time asymptotic values are given by scaling:
$\left\langle r^2(t) \right\rangle =
e^{2 l^*} \left\langle r^2(t(l^*), l^*) \right\rangle$ 
and $c_A(t) = e^{-2 l^*} c_A(t(l^*), l^*)$.
This procedure gives
\begin{equation}
c_A(t) \sim \left( \frac{1}{2 \pi D y} + \frac{1}{\lambda_0} \right) \frac{1}{t},
~~~(\gamma =0)
\label{13}
\end{equation}
where we recognize the mean field result with  effective reaction rate
$1/k^* = 1/(2 \pi D y) + 1/\lambda_0$.  The mean square displacement
is given by $\langle r^2(t) \rangle \sim 4 D t (t/t_0)^{y/(4 - y)}$,
which is an exact law \cite{Bouchaud3} in the absence of
reaction.  As expected,
turbulence leads to the well-mixed, mean-field result for the
concentration decay, with an effective reaction rate
$k^* < \lambda_0$.  We have determined the effective reaction rate
as an expansion in the parameter $y$, which measures the degree of
mixing of the fluid.  For simplicity,
here, and below, we have used used only the fixed point value of $g$
when integrating the flow equation for $\lambda(l)$.

What will happen in the presence of a quenched, random potential?
We have seen that fluid streamlines increase the mixing of the
reactants.  We have previously shown that a quenched, random
potential in the absence of turbulence leads to a slowing down of
the reaction \cite{Deem1}.  This occurs because the reaction becomes
diffusion-limited at long times, and the random potential leads to
sub-diffusion.  We might, therefore, expect that the
reactant concentration 
for $\gamma \ne 0$ will be lower than that for
the case of $\gamma =0$ for arbitrary
$\sigma$.  In fact, a subtle trapping effect due to the potential
leads to increased decay rates, above that for $\gamma =0$, for some
intermediate values of $\gamma$.

\begin{figure}[t]
\centering
\leavevmode
\psfig{file=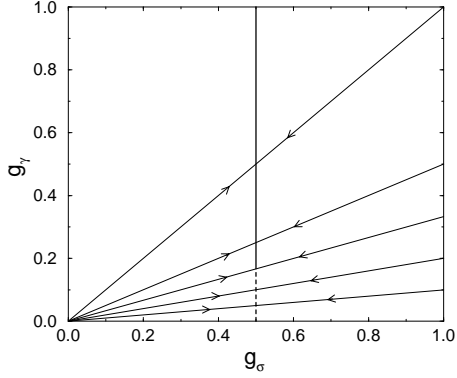,height=2in,angle=-90}
\caption[]
{\label{fig1}
The flow diagram for the dimensionless couplings in the $A+A \to \emptyset$
reaction.  The fixed line is shown in bold for $y=1$.  The fixed point
reaction rate is finite on the solid part of the fixed
line and vanishes on the dashed part of the fixed line.
}
\end{figure}

The flow equations that result in the presence of the random potential
are
\begin{eqnarray}
\frac{d \ln n_0}{d l} &=& 2
\nonumber \\
\frac{d \ln \lambda}{d l} &=& - 
   \frac{\lambda}{4 \pi D} +3 g_\gamma - g_\sigma 
\nonumber \\
\frac{d \ln g_\gamma}{d l} &=& y  - 2 g_\sigma
\nonumber \\
\frac{d \ln g_\sigma}{d l} &=& y  - 2 g_\sigma
\ ,
\label{14}
\end{eqnarray}
where the dimensionless coupling constants are given by
$g_\gamma = \gamma \beta^2 \Lambda^{-y} / (4 \pi)$ and
$g_\sigma = \sigma \beta^2 \Lambda^{-y} / (4 \pi)$.
The dynamical exponent is given by
\begin{equation}
z = 2  +  g_\gamma- g_\sigma  \ .
\label{15}
\end{equation}
These flow equations lead to fixed points for the couplings
$g_\sigma^* = y/2$ and $g_\gamma^* = (\gamma / \sigma) y/2$.
The flow diagram for the couplings is shown in Figure \ref{fig1}.
Including higher order diagrams in the flow equation leads
to a bending of the fixed line,
 more strongly for larger $\gamma/\sigma$,
 with the coupling flows
remaining linear \cite{Honkonen3}.

\begin{figure}[t]
\centering
\leavevmode
\psfig{file=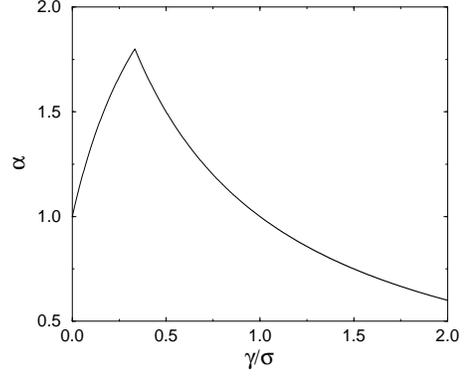,height=2in,angle=-90}
\caption[]
{\label{fig2}
The decay exponent for the
$A+A \to \emptyset$ reaction: $c_A(t) \sim ({\rm const}) t^{-\alpha}$.
The figure is shown for 
$y = 8/3$, which mimics isotropic turbulence.
}
\end{figure}

The matching to determine the asymptotic concentrations leads to
two regimes.
For weak potential disorder with
$3 \gamma < \sigma$, there is no finite
fixed point value for $\lambda(l)$, and we have
\begin{eqnarray}
c_A(t) &\sim& \left[ \frac{1}{4 \pi D (g_\sigma^* - 3 g_\gamma^*)}
 + \frac{1}{\lambda_0} \right] \frac{1}{t}
\nonumber \\
&&\times
\left( \frac{t}{t_0} \right)^{-
 2 g_\gamma^*/(2 + g_\gamma^* - g_\sigma^*)} , ~~~(3 \gamma < \sigma) \ .
\label{16} 
\end{eqnarray}
For strong potential disorder with
$3 \gamma > \sigma$, there is a finite fixed point value $\lambda^* = 
4 \pi D (3 g_\gamma^* - g_\sigma^*)$, and
we have
\begin{equation}
c_A(t) \sim  \frac{1}{\lambda^* t}
\left(\frac{t}{t_0}\right)^{
(g_\gamma^* - g_\sigma^*)/(2 + g_\gamma^* - g_\sigma^*)} , 
~~~(3 \gamma > \sigma) \ .
\label{17} 
\end{equation}
The maximum rate of decay occurs for $3 \gamma = \sigma$, in which case
we have
\begin{equation}
c_A(t) \sim  \frac{\ln (t/t_0)}{8 \pi (1 -y/6) D t}
t^{-y/(6-y)} 
, ~~~(3 \gamma = \sigma) \ .
\label{18} 
\end{equation}
In all cases, the mean square displacement is given by
$\langle r^2(t) \rangle \sim 4 D t (t/t_0)^{(g_\sigma^* - g_\gamma^*)/(2 - 
g_\sigma^* + g_\gamma^*)}$,

Examining the asymptotic decay laws (\ref{16})-(\ref{18}), we see that a small
amount of potential disorder added to the turbulent fluid mixing leads
to an {\em increased} rate of reaction.  As the potential disorder is
increased, eventually the rate of reaction decreases.  The exponent of
the concentration decay is shown in Figure \ref{fig2}.
  This result is
rigorously valid for small $y$ and finite values of
$\gamma$ and $\sigma$.  For finite values of $y$, we expect 
qualitatively  similar behavior.

How can potential disorder, which tends to slow down the 
diffusing reactants, lead to an increased rate of reaction?
The potential disorder creates regions of low energy, which
tend to attract reactants.  The local density of reactants in 
these deep energy wells is significantly higher than the average
density.  The reaction rate in these regions, therefore, is higher
than would be predicted by mean-field theory based upon the
average density.  The turbulent mixing
flows
continuously replenish the reactants in these wells as the
reaction occurs.  In this way, a reaction rate significantly
higher than that 
for a perfectly well-mixed system arises.

\section{The $A+B \to \emptyset$ Reaction}
We now turn to the $A+B \to \emptyset$ reaction.  In this case,
the $A$ and $B$ reactants are attracted to different regions of
space by the external potential, and so there is no mechanism for
super-fast reaction.

The flow equations for this case are 
\begin{eqnarray}
\frac{d \ln n_0}{d l} &=& 2
\nonumber \\
\frac{d \ln \lambda}{d l} &=& -
   \frac{\lambda}{4 \pi D} -\left( g_\gamma + g_\sigma \right)
\nonumber \\
\frac{d \ln g_\gamma}{d l} &=& y  - 2 g_\sigma
\nonumber \\
\frac{d \ln g_\sigma}{d l} &=& y  - 2 g_\sigma \ .
\label{19}
\end{eqnarray}
The dynamical exponent is given by
\begin{equation}
z = 2  +  g_\gamma- g_\sigma  \ .
\label{19a}
\end{equation}

We first consider the case of no external potential.  If, in addition,
there is no turbulent flow, the $A$ and $B$ reactants segregate into
distinct regions in space.  This segregation leads to a severely
diffusion-limited reaction at long times.  The concentration 
decays as $c_A(t) = c_B(t) \sim [n_0 / (8 \pi^2 D t)]^{1/2}$ \cite{Deem2}.
Allowing for turbulent mixing, we expect the reaction to become more
well-mixed, with less segregation and faster reaction.  In fact,
there will be a transition to a region that is reaction limited for
strong enough mixing flows.

We perform the matching to determine the asymptotic decay.
In the transport-limited regime,  we use
$c_{\rm A}(t(l^*),l^*) = 
[ n_0(l^*)/(8 \pi^2 D t(l^*)) ]^{1/2}$.
In the reaction-limited regime, we use
$c_A(t(l^*), l^*) = 1/[1/n_0(l^*) + \lambda(l^*) t(l^*)]$, as
before.

\begin{figure}[t]
\centering
\leavevmode
\psfig{file=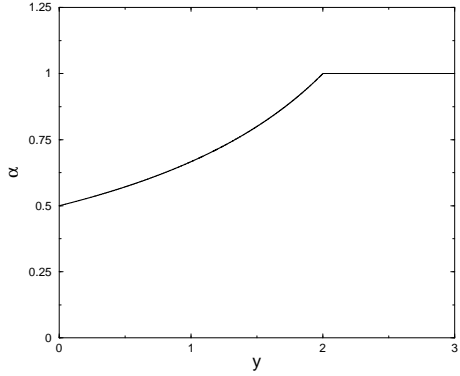,height=2in,angle=-90}
\caption[]
{\label{fig3}
The decay exponent for the
$A+B \to \emptyset$ reaction: $c_A(t) \sim ({\rm const}) t^{-\alpha}$.
The figure is shown for $\gamma = 0$.
}
\end{figure}

For weak fluid mixing, the reaction will be in the transport-limited
regime, whereas for strong fluid mixing, the reaction will be in the
reaction-limited regime.  Specifically, for weak mixing we have
\begin{equation}
c_A(t) \sim \left( \frac{n_0}{8 \pi^2 D t} \right)^{1/2}
\left( \frac{t}{t_0}\right)^{-y/(8-2y)}, ~~~(y < 2, \gamma=0) \ .
\label{20}
\end{equation}
For strong mixing, we have
\begin{equation}
c_A(t) \sim \left( \frac{1}{2 \pi D y} + \frac{1}{\lambda_0} \right) \frac{1}{t},
~~~(y > 2, \gamma=0) \ .
\label{21}
\end{equation}
The exponent of this decay is shown in Figure \ref{fig3}.
Note that for isotropic turbulence ($y = 8/3$), the reaction
is always in the reaction-limited regime in two dimensions.

\begin{figure}[t]
\centering
\leavevmode
\psfig{file=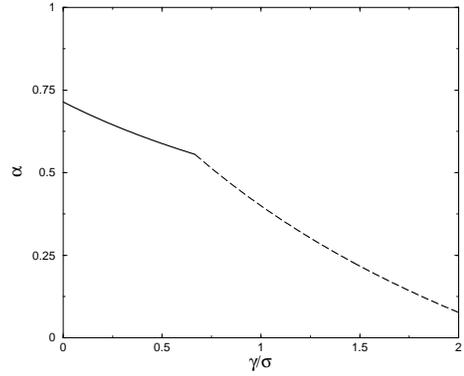,height=2in,angle=-90}
\caption[]
{\label{fig4}
The decay exponent for the
$A+B \to \emptyset$ reaction: $c_A(t) \sim ({\rm const}) t^{-\alpha}$.
The figure is shown 
for $y = 1.2$.  The reaction is transport-limited on
the solid curve and reaction-limited on the dashed curve.
The curve is strictly valid only for
small $\gamma/\sigma$.
}
\end{figure}

Potential disorder will slow down the reaction, both because
reactants are attracted to different regions of space and because
the transport of
reactants to each other is slowed. 
Our one-loop flow equations predict that in the transport-limited regime
\begin{eqnarray}
c_A(t) \sim \left( \frac{n_0}{8 \pi^2 D t} \right)^{1/2} &&
\left( \frac{t}{t_0}\right)^{(g_\gamma^* - g_\sigma^*)/(4 - 2
g_\sigma^* + 2 g_\gamma^*)},
\nonumber \\
&&(g_\sigma^*+g_\gamma^* < 1 ) \ ,
\label{22}
\end{eqnarray}
where, as before, we have
$g_\sigma^* = y/2$ and $g_\gamma^* = (\gamma / \sigma) y/2$.
In the reaction-limited regime, we have
\begin{eqnarray}
c_A(t) &\sim& \left[ \frac{1}{4 \pi D (g_\sigma^* + g_\gamma^*)}
 + \frac{1}{\lambda_0} \right] \frac{1}{t}
\nonumber \\
&&\times
\left( \frac{t}{t_0}\right)^{2 g_\gamma^* / (2 - g_\sigma^* + g_\gamma^*)}
, ~~~(1<g_\sigma^*+ g_\gamma^* < 2) \ .
\label{23}
\end{eqnarray}
The exponent of this decay is shown in Figure \ref{fig4}.
The exponent is valid for arbitrary $y$ and small
$\gamma/\sigma$.  The effective reaction rate in Eq.\ (\ref{23}), however, may
contain corrections higher order in $y$.

\section{Conclusion}
Experiments to test our predictions for isotropic turbulence, Eqs.\ 
(\ref{13}) and (\ref{21}), would be relatively simple to perform.
The behavior of the prefactor would be the quantity of interest.
Reaction conditions of the type that we consider could be realized in
reactions between ionic species confined to
two-dimensional fluid films that are surrounded by spatially-addressable
electrodes or  media with ionic disorder
that is {\em not} equilibrated.  The electrodes or disordered media
are necessary to generate a potential with the required, singular
correlation function.  
The required isotropic turbulence can be generated in the
standard fashion.
Fluid flows less strong
than isotropic turbulence ($y < 8/3$) could be observed in regions of
developing turbulence.  

A recent experiment by Paireau and Tabeling has
seen an enhancement of the effective reaction rate between ions
in a chaotically-mixed, 
two-dimensional,
fluid with attractors \cite{Tabeling}.
  In this
experiment only the prefactor to the reactivity was enhanced.
The decay exponent remained at unity
because the disorder was technically irrelevant.
We are unaware of experiments, to date, that
can test our predictions for technically
relevant disorder, Eqs.\ (\ref{16})-(\ref{18}) and
 (\ref{22})-(\ref{23}).

\section*{Acknowledgment}
This research was supported by the National Science Foundation
through grants CHE--9705165 and CTS--9702403.

\bibliography{react4}

\end{document}